\documentclass [12pt,a4paper,leqno]{article}

\title{Static Cylindrically Symmetric Solution}
\author{Mihaela Time}
\date{6th~December~1998}

\begin{document}

\maketitle

\begin{abstract}
Here is described a static axisymmetric solution with an additional cylindrical
symmetry for which the matter consists in a cosmological and a dust term.
\end{abstract}

\section{The field equations}
Stationary gravitational fields are characterized by the existence of a timelike
Killing vector field $\xi$. So, in a stationary space-time $(M,\,g)$ we can have
a global causal structure, i.e. we can introduce an adapted coordinate system
$(x^a) = (x^{\alpha},\,t)$, $\xi = \frac{\partial}{\partial t}$ so that the
metric $g_{ab}$ is independent of $t$,
\begin{equation} \label{eq:stat}
 \mathrm{d}s^2 = h_{\alpha\beta}\mathrm{d}x^{\alpha}\mathrm{d}x^{\beta} +
F{(\mathrm{d}t + A_{\alpha}\mathrm{d}x^{\alpha})}^2, \qquad F \equiv \xi_a\xi^a
< 0.
\end{equation}
The timelike unitary vector field $h^0 \equiv {(-F)}^{-{1\over 2}}\xi$, which is
globally defined on $M$, indicates not only the time-orientation in every point
$p \in M$, but also gives a global time coordinate $t$ on $M$.

Stationarity (i.e. time translation symmetry) means that there exists a
1-dimensional group $G_1$ of isometries $\phi_t$ whose orbits are timelike
curves parametrized by $t$. The smooth surjective submersion
$$\pi': M \longrightarrow \mathcal{S}_3, \quad p \in M \longmapsto \pi'(p) =
\phi_t(p), \enspace \phi_0(p) = p, \enspace \dot \phi_p(t) = \xi(\phi_p(t))$$
over the 3-dimensional differentiable factor manifold $\mathcal{S}_3$, whose
elements are the orbits of $G_1$ determines a fibration on $M$; the metric
$g_{ab}$ is constant on the fibres of $\pi'$ (integral curves of $\xi$) and
there is a 1:1 correspondence between tensors $T = (T_{b\ldots}^{a\ldots})$on
$M$ satisfying $$\xi^aT_{a\ldots}^{b\ldots} = \xi_bT_{a\ldots}^{b\ldots} = 0
\textrm{\ and\ } \pounds_{\xi}T = 0$$ and the tensors on $\mathcal{S}_3$. So
using the 3-projection formalisme (by Geroch(1971)) developed on
$\mathcal{S}_3$ (see \cite{mc:gravitation}), the Einstein's field equations
\begin{equation}
 R_{ab} - {1\over 2} R g_{ab} + \Lambda g_{ab} = \kappa T_{ab},
\end{equation}
will take the following simplified form for stationary fields:
\begin{equation} \label{eq:einstein}
\left \{
\begin{array}{l}
\displaystyle R^{(3)}_{ab}={1\over 2}F^{-2}(\frac{\partial F}{\partial
x^a}\frac{\partial F}{\partial x^b}+\omega_{a}\omega_{b})
+\kappa(h^c_ah^d_b-F^{-2}\tilde{h}_{ab}\xi^a\xi^b)(T_{cd}-{1\over 2}Tg_{cd});

\vspace{6pt} \\ \vspace{6pt}

\displaystyle F^{\parallel a}_{,a}= F^{-1}\tilde{h}_{ab}(\frac{\partial
F}{\partial x^a}\frac{\partial F}{\partial x^b}-\omega_{a}
\omega_{b})-2\kappa F^{-1}\xi^a\xi^b(T_{ab}-{1\over 2}Tg_{ab});\\
\displaystyle \omega^{\parallel a}_a=2F^{-1}\tilde{h}_{ab}\frac{\partial
F}{\partial x^a}\omega_b

\vspace{6pt} \\ \vspace{6pt}

\displaystyle F \epsilon^{abc} \omega_{c,b} = 2 \kappa h_b^a T_c^b \xi^c
\end{array}
\right .
\end{equation}

Here, ``$\parallel$'' denotes the covariant derivative associated with the
conformal metric tensor $\tilde{h}_{ab} = -F h_{ab}$ on $\mathcal{S}_3$
($h_{ab} = g_{ab} + h^0_a h^0_b$ is the projection tensor), $\epsilon^{abc} =
\epsilon^{dabc}h^0_d$ and $\omega^a = {1\over 2}\epsilon^{abcd} \xi_{b} \xi_{c;d}
\neq 0$ is the rotational vector
($\omega^a\xi_a=0,\enspace \pounds_\xi\omega=0$).

The circularity theorem (due to Kundt) states that an (axisymmetric) metric can
be written \cite{hs:stationary} in the (2+2)-split if and only if the conditions
\begin{equation}
(\eta^{[a}\xi^b\xi^{c;d]})_{;e}=0=(\xi^{[a}\eta^b\eta^{c;d]})_{;e}
\end{equation}
are satisfied \footnote{Here I use the convention: round brackets denote symmetrization
and square brackets antisymmetrization.}.

The existence of the orthogonal 2-surfaces is possible for dust solutions,
provided that the 4-velocity of dust satisfies the condition:
\begin{eqnarray}
u_{[a}\xi_b\eta_{c]}=0,\quad u^a=(-H)^{-{1\over 2}}(\xi^a+\Omega\eta^a)=
(-H)^{-{1\over 2}}l^i\xi^a_i,\quad\textrm{where}\\
l^i\equiv(1,\Omega),\quad H=\gamma_{ij}l^il^j,\quad \gamma_{ij}\equiv
\xi^a_i\xi_{aj},\quad i,j=1,2,\quad \xi_1=\xi;\ \xi_2=\eta \nonumber
\end{eqnarray}
i.e., the trajectories of the dust lie on the transitivity surfaces of the group
generated by the Killing vectors $\xi$, $\eta$. Here $\Omega$ is the angular velocity
of the matter with respect to infinity.

Using an adapted coordinate system, the metric (\ref{eq:stat}) can be written in
the following form:
\begin{equation} \label{eq:dust}
\mathrm{d}s^2=e^{-2U}[e^{2V}(\mathrm{d}r
^2+\mathrm{d}z^2)+W^2\mathrm{d}\varphi ^2]-
e^{2U}(\mathrm{d}t+A\mathrm{d}\varphi )^2
\end{equation}
where the functions \footnote{The function $W$ is defined invariantly as
$W^2 \equiv -2\xi_{[a}\eta_{b]}\xi^a\eta^b$.} $U$, $V$, $W$ and $A$
depend only on the coordinates $(r,z)$; these coordinates are also conformal
flat coordinates on the 2-surface $S_2$ orthogonal to 2-surface $T_2$ of the
commuting Killing vectors $\xi = \partial_t$ and $\eta = \partial_\varphi$.

If we identify the 4-velocity of the dust $u^a$ with timelike Killing vector
$\xi^a = \partial_t = (0, 0, 0, 1)$ then (\ref{eq:dust}) represents a co-moving
system $(x^1=r,\ x^2=z,\ x^3=\varphi,\ x^0=t)$ with dust,
$u_a = \xi_a = (0,\, 0,\, -e^{2U}A,\, -e^{2U})$ and
\begin{equation} \label{eq:metric}
\left \{
\begin{array}{l}
g_{11}=g_{22}=e^{-2U+2V}=h_{11}=h_{22},\\
g_{33}=e^{-2U}W^2-e^{2V}A^{2}=h_{33},\quad g_{00}=\xi_0=-e^{2U}=F,\\
g_{03}=\xi_3=-e^{2U}A,\quad g_{13}=g_{23}=g_{10}=g_{20}=0
\end{array}
\right .
\end{equation}

The energy-momentum tensor $T_{ab}$ has the
form as:
\begin{equation}\label{eq:momentum}
\kappa T_{ab}=-\Lambda g_{ab}+\mu u_a u_b,\quad \mu>0,
\Lambda =\textrm{const.}>0.
\end{equation}

The conservation law, $T_{;b}^{ab}=0$ implies $U_{,a}=0$ and is a consequence
of the field equations, being used in place of one of the equations (\ref{eq:einstein}).

If $\Omega=\textrm{const.}$, i.e., for rigid rotation we have
\begin{equation}
\sigma=0=\theta\Leftrightarrow u_{(a;b)}+u_{(a}\dot{u}_{b)}=0
\end{equation}
and using the simplification $U = 0$ in (\ref{eq:metric}), then the matter current is
geodesic ($\dot{u}_a=u_{a;b}u^b=0$), without shear
($\sigma_{ab}=u_{(a;b)}+\dot{u}_{(a}u_{b)}-{1\over 3}\theta h_{ab}=0$) and
without expansion ($\theta=u_{;a}^a=0$); this implies that $A_{,r}=0=A_{,z}$.
Moreover the dust is without rotation, $omega_{ab}=u_{[a;b]}+\dot{u}_{[a}u_{b]}=0$.

If we take $A=0, U=0$ then
\begin{equation} \label{eq:metric2}
\left \{
\begin{array}{l}
g_{11}=g_{22}=e^{2V}=h_{11}=h_{22},\\
g_{33}=W^2=h_{33},\quad g_{00}=\xi_0=-1=F,\\
g_{03}=\xi_3=0=g_{13}=g_{23}=g_{10}=g_{20}
\end{array}
\right .
\end{equation}
and the field equations (\ref{eq:einstein}) will have the following simplified form:
\begin{equation}\label{eq:einstein3}
\left \{
\begin{array}{l}
\displaystyle
{1\over 2}\Bigl(\frac{\partial^2W}{\partial r^2}+\frac{\partial^2W}{\partial z^2}\Bigr)
=-\Lambda We^{2V}
\vspace{6pt} \\ \vspace{6pt}
\displaystyle
(\mu-2\Lambda)e^{2V} = 0
\vspace{6pt} \\ \vspace{6pt}
\displaystyle
{1\over 2}\Bigl(\frac{\partial^2W}{\partial r^2}+\frac{\partial^2W}{\partial z^2}\Bigr)=
\frac{\partial W}{\partial r}\frac{\partial V}{\partial r}-
\frac{\partial W}{\partial z}\frac{\partial V}{\partial z}
\vspace{6pt} \\ \vspace{6pt}
\displaystyle
\frac{\partial W}{\partial r}\frac{\partial V}{\partial z}+
\frac{\partial W}{\partial z}\frac{\partial V}{\partial r}=0
\vspace{6pt} \\ \vspace{6pt}
\displaystyle
\frac{\partial^2V}{\partial r^2}+\frac{\partial^2V}{\partial z^2}= -\Lambda e^{2V}
\end{array}
\right .
\end{equation}
where $W$ and $V$ are functions of $x^1=r$ and $x^2=z$.

Finally, taking into account the third symmetry ($\zeta = \partial_z$ is the
spacelike Killing vector field parallel with the rotation axis $z$) as a special case
of the stationary axisymmetric (with $\xi$ and $\eta$) solution (\ref{eq:einstein3})
we will obtain:
\begin{equation}
\mathrm{d}s^2=e^{2V(r)}(\mathrm{d}r ^2+\mathrm{d}z^2)+
W^{2}(r)\mathrm{d}\varphi ^2 - \mathrm{d}t^2
\end{equation}
where $W(r)$ and $V(r)$ will be determined from (\ref{eq:einstein3}).

\section{The solution}
Because of the third symmetry $W$ and $V$ depend only on $r$ and the field
equations are reduced to:
\begin{equation} \label{eq:sysfinal}
\left \{
\begin{array}{l}
W''=-2\Lambda We^{2V} \\
W''=2W'V' \\
V''+\Lambda e^{2V}=0
\end{array}
\right .
\end{equation}
where ${\partial \over \partial r}='$ and $\mu = 2\Lambda = \mathrm{const.}$.

The last equation in the (\ref{eq:sysfinal}) system is a differential equation of second
order which implies:
\begin{equation} \label{eq:sol1}
 r=\int \frac{1}{\sqrt{\gamma-\Lambda e^{2V}}}\, \mathrm{d}V + \alpha
\end{equation}
where $\gamma$ and $\alpha$ are constants.

Integrating in (\ref{eq:sol1}) we find:
\begin{equation}
 r={1\over 2\lambda}\mathrm{ln}\Biggl| \frac{\sqrt{\gamma-\Lambda e^{2V}}-\lambda}
{\sqrt{\gamma-\Lambda e^{2V}}+\lambda} \Biggr| +\nu
\end{equation}

Finally we obtain:
\begin{equation}
 V(r)={1\over 2}\mathrm{ln} \Bigl( \gamma-\lambda^2 \Biggl( \frac{1-e^{2\lambda (r-\nu)}}
{1+e^{2\lambda (r-\nu)}} \Biggr)^2 \Bigr) - \mathrm{ln}\sqrt{\Lambda}
\end{equation}

The first two equations in (\ref{eq:sysfinal}) reduce to
\begin{equation}
 {W'\over W}=-\Lambda {e^{2V}\over V'}
\end{equation}
which becomes:
\begin{equation}
 \mathrm{ln}W=\lambda \int \frac{1+e^{2\lambda (r-\nu)}}{1-e^{2\lambda (r-\nu)}}
\, \mathrm{d}r - {\gamma\over \lambda^2}
\int \frac{1-e^{2\lambda (r-\nu)}}{1+e^{2\lambda (r-\nu)}}\, \mathrm{d}r
\end{equation}

Integrating both integrals we are lead to:
\begin{equation}
W(r)=\frac{1-e^{2\lambda (r-\nu)}}{(1-e^{2\lambda (r-\nu)})^{\gamma \over \lambda^3}}\,
e^{2\lambda (r-\nu)({1\over 2}-{\gamma \over 2\lambda^3})}
\end{equation}

where $\gamma$, $\alpha$, $\lambda$ and $\nu$ are constants.

The resulting solution is static cylindrically symmetric and conformally flat and the matter
consists in a cosmological and dust term with $\mu=2\Lambda=\mathrm{const.}$
positive.

{\bf Remark:} Further asking the regularity condition on the axis of rotation to be satisfied, i.e.:
\begin{equation}
\lim_{r \to 0} \frac{\frac{1-e^{2\lambda (r-\nu)}}{(1-e^{2\lambda (r-\nu)})^{\gamma \over \lambda^3}}\,
e^{2\lambda (r-\nu)({1\over 2}-{\gamma \over 2\lambda^3})}}{r\sqrt{\lambda^2(\frac{1+e^{2\lambda(r-\nu)}}
{1-e^{2\lambda(r-\nu)}})^2-\gamma^2}}=1
\end{equation}
we find out the regular Einstein's static universe (see \cite{mc:gravitation}) solution in cylindrically
coordinates (with $\mu=2\Lambda={1\over 2K^2}=\mathrm{const.}$).

\end{document}